\documentclass[a4paper, useAMS, usenatbib]{mn2e}
 \usepackage{amsmath, amssymb, mathrsfs, dcolumn, aas_macros,bm, url, dcolumn, graphicx}
\newcommand{\bq}{\begin{equation}}
\newcommand{\eq}{\end{equation}\noindent}
\newcommand{\bqa}{\begin{eqnarray}}
\newcommand{\eqa}{\end{eqnarray}\noindent}
\newcommand{\bsub}{\begin{subequations}}
\newcommand{\esub}{\end{subequations}}
\newcommand{\ba}{\begin{array}}
\newcommand{\ea}{\end{array}\noindent}
\newcommand{\half}{\tfrac{1}{2}}
\newcommand{\brak}[1]{\left[ #1 \right]}
\renewcommand{\vec}{\mathbf}

\begin{document}

\title[Scattering in relativistic an-isotropic plasmas]{Scattering of magnetosonic waves in a relativistic and an-isotropic magnetised plasma}

\author[J.~Moortgat and J.~Kuijpers]{Joachim Moortgat$^{1}$\thanks{Email: moortgat@astro.ru.nl. Current address: Department of Physics \& Astronomy, University of Rochester, Bausch \& Lomb Hall, P.O. Box 270171, 600 Wilson Boulevard, Rochester, NY 14627-0171.} and Jan Kuijpers$^{1}$\\
$^{1}$ Department of Astrophysics, {\sc imapp}, Radboud University, PO Box 9010, 6500 GL Nijmegen, The Netherlands.}

\maketitle

\begin{abstract}
Gravitational waves ({\sc gw}) propagating through a magnetised plasma excite low-frequency magnetohydrodynamic ({\sc mhd}) waves. 
In this paper we investigate whether these waves can produce observable radio emission at higher frequencies by scattering on an an-isotropic intrinsically relativistic distribution of electrons and positrons in the force-free wind surrounding a double neutron star binary merger. 
The relativistic particle distribution is assumed to be strictly along the magnetic field lines, while the magneto-plasma streams out at a relativistic speed from the neutron stars. In the case of Compton scattering of an incident {\sc mhd} wave transverse to the magnetic field, we find that the probability of scattering to both a transverse $x$-mode and a quasi-longitudinal Langmuir-$o$ mode is suppressed when the frequency of the incident wave is below the local relativistic gyro-frequency, i.e.~when the magnetic field is very strong. 
\end{abstract}

\section{Introduction}
This decade is expected to witness the historical first direct detection of gravitational waves ({\sc gw}) with detectors such as the Advanced Laser Interferometer Gravitational Wave Observatory ({\sc ligo}\footnote{{\sc ligo} website: {\tt http://www.ligo.caltech.edu/}}) and the Laser Interferometer Space Antenna ({\sc lisa}\footnote{{\sc lisa} website: {\tt http://lisa.jpl.nasa.gov/}}). The former is designed to detect gravitational waves from compact sources such as merging neutron star binaries in the kHz regime, whereas the latter will detect the lower frequency ($<1$~Hz) gravitational waves from thousands of compact white dwarf binaries \citep{nelemans} as well as the spiral-in signal of super-massive black holes, throughout the visible universe (see \cite{schutz} or \cite{kokkotas} for a review of likely sources for {\sc ligo} and {\sc lisa}). 

Gravitational waves are emitted by highly energetic events but occur at relatively large distances and the signal that reaches Earth is exceedingly weak. To extract the {\sc gw} from the noisy signal, some theoretical knowledge of the expected waveforms is essential. To confirm the detection of a {\sc gw} burst, any additional electromagnetic signature of the event would be extremely useful.

It so happens that many of the possible {\sc gw} sources are embedded in a strong magnetic field. Examples are rapidly spinning neutron stars which have a small oblateness and precess, accrete, or are unstable to the excitation of the $r$-mode, supernova core collapses and bounces, newly born `boiling' and oscillating neutron stars, magnetars with crust fracturing ($\sim 20$~Hz), and coalescing compact binaries in which at least one component is a magnetic neutron star.

In the last case, numerical models predict that maximum {\sc gw} luminosities of the order of $10^{48}$~W \citep{ibrahim} are released into a wound-up magnetic field of up to $10^{8}$ -- $10^{12}$~T. 
We have investigated in previous papers (\cite{moortgatI, moortgatII, moortgatIII} referred to from hereon as Paper I--III) whether these extreme space-time distortions 
perturb the ambient magnetic field sufficiently to produce an observable electromagnetic counterpart of the {\sc gw} burst. We have found that a {\sc gw} can couple linearly at the same frequency to all three low-frequency plasma eigenmodes. The shear Alfv{\'e}n mode and the transverse slow and fast  magnetosonic waves ({\sc msw}) are excited depending on the polarisation of the gravitational waves with respect to the ambient magnetic field orientation, but only the fast magneto-acoustic mode can interact coherently with the {\sc gw} a over longer time (distance) and is therefore the most interesting.

The {\sc msw} can not escape directly as observable radiation. In Paper III we pointed out that the {\sc msw} encounters relativistic electrons and positrons with Lorentz factors different from the bulk Lorentz factor of the particles carrying the {\sc msw}, for instance due to a tail in the distribution function or simply due to inhomogeneities in the plasma such as beams. For typical Lorentz factors of a few hundreds, the {\sc msw} would than be inverse Compton ({\sc ic}) scattered to low-frequency radio emission in the {\sc lofar} (the LOw Frequency Radio Array\footnote{{\sc lofar} website: {\tt http://www.lofar.org/}}) band. In a first estimate, we approximated the scattering cross-section with the Thomson cross-section and predicted a flux that would be easily detectable by {\sc lofar} for a double neutron star merger at a distance of $\leq 1$~Gpc.
Clearly, the indirect detection of gravitational waves with a radio array would be of extreme interest. 

In the present paper we study in more detail the scattering process of a fast {\sc msw} into escaping electromagnetic waves in a strongly magnetised an-isotropic relativistic plasma.

The paper is set up as follows. In Section~\ref{eq::recap} we recapitulate the intuitive picture presented in Paper II which underlies the excitation of {\sc mhd} waves by a gravitational wave propagating through a magnetised plasma. To investigate the subsequent scattering of these {\sc mhd} waves we need a covariant relativistic theory for plasma dynamics. Such a theory was developed largely in \citet{gedalin98, melrose99, melrosegedalin99,melrose01, luomelrose, melrose02} and unpublished work in \citep{melrosebook} to study emission processes in pulsar magnetospheres.  
We apply this theory to the force-free wind well outside the effective light cylinder of the merging binary. In particular, we consider a plasma which --in the frame comoving with the wind-- has a momentum dispersion along the magnetic field only, and construct the covariant equations for the {\sc gw}-excited current, the linear and non-linear response to that current and the scattering probability in Sections~\ref{sec::emission}--\ref{sec::responses}.
In Section~\ref{sec::scattering} we apply these general results to the geometry of an incident transverse magnetosonic wave that scatters on electrons and positrons with a $1$-dimensional relativistic dispersion along the magnetic field and calculate the probability that either transverse or longitudinal waves are emitted in the direction of the observer. A numerical analysis is 
presented in Section~\ref{sec::discussion}, where we calculate the energy radiated into scattered radio waves. The conclusions follow in Section~\ref{sec::conclusions}.

Standard geometrised units are used throughout this discussion ($G=c=1$), Greek indices are used for time-space components ($\mu = 0 \ldots 3$) and Latin indices for spatial only components ($i = 0 \ldots 3$).

%%%%%%%%%%%%%%%%%%%%%%%%%%%%%%%
\section{GW $\Rightarrow$ magneto-acoustic wave}\label{eq::recap}
\begin{figure}
\resizebox{\hsize}{!}{\includegraphics{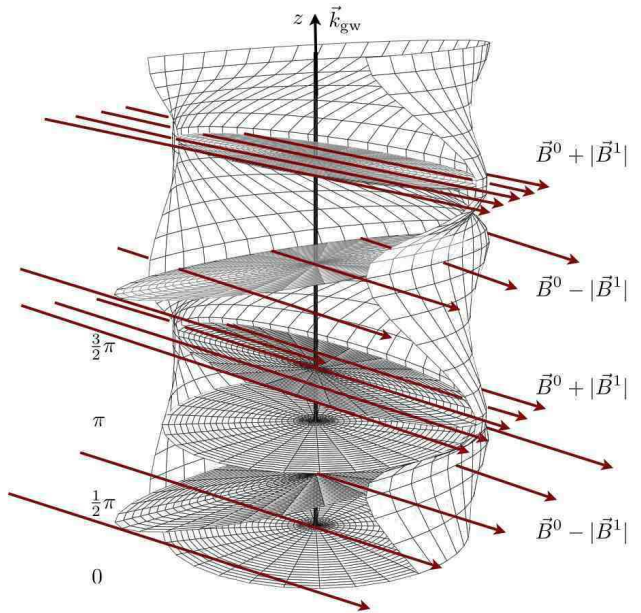}}
\caption{\label{fig::effect}Deformation of the world-sheet of a ring of test particles by the tidal force of a passing $+$ polarised {\sc gw} (with $\vec{k}_{\mathrm{gw}}\bot \vec{B}^{0}$). The magnetic field is stretched and compressed and excites a magneto-compressional wave.}
\end{figure}

%%%%%%%%%%%%%%%%%%%%%%%%%%%%%%%
In a pulsar environment plasma flows out {\em along} the open field lines and develops into a force-free wind outside the light-cylinder in which the toroidal component of the magnetic field quickly dominates the poloidal one (see Section~\ref{sec::discussion} and Fig.~\ref{fig::gwbdance}). The magnetic field is therefore essentially perpendicular to the radial propagation of the wind as in Fig.~\ref{fig::effect}.
In this section we discuss what happens when the two neutron stars at the origin of the system merge and release a large amount of {\sc gw} into the wind.

\subsection{Gravitational wave tidal field}
%%%%%%%%%%%%
%%%%%%%%%%%%
The driving force exerted by a {\sc gw} on test particles is described by Einstein's field equations or alternatively by the general relativistic equations for {\em geodesic deviations} describing tidal accelerations (see for instance \cite{stewart}).
Integrating those differential equations twice results in the well-known spatial deviations of test masses in an interferometer detector such as {\sc ligo}: 
%.....................................................%
\begin{equation}\label{eq::detector}
\delta x = \frac{1}{2} (h_+ x_0 + h_\times y_0)\ , 
\qquad
\delta y = \frac{1}{2} (h_\times x_0 - h_+ y_0)\ ,
\end{equation}
%.....................................................%
where $x_{0}$ and $y_{0}$ are the initial separations of test-masses along the $x$- and $y$-axis in an L-shaped interferometer, $\delta x$ and $\delta y$ are the excursions causes by the passing {\sc gw} and $h_{+}$ and $h_{\times}$ refer to the two possible polarisations of the {\sc gw}.

\subsection{Interaction with magnetic field}
In the ideal {\sc mhd} approximation the magnetic field lines are frozen into the plasma, and the magnetic field will exhibit the same excursions as the test particles. The action of a {\sc gw} propagating in the $z$ direction is only in the $x-y$ plane, and when we choose the orientation of the ambient magnetic field in the $x-z$ plane, a {\sc gw} which is $+$ polarised with respect to these axes (so $h_{\times}=0$) excites a growing magnetic field perturbation with:
%.....................................................%
\begin{equation}\label{eq::bx}
\delta B_x \propto \frac{1}{2} h_+ B_x^0\ , 
\qquad 
\delta B_y \propto \frac{1}{2} h_+ B_y^0 = 0\ ,
\end{equation}
%.....................................................%
The resulting interaction is illustrated in Fig.~\ref{fig::effect} where we show the world-sheet of a ring of test particles in the tidal field of a passing {\sc gw}, which we choose to propagate in the $z$-direction (and $\vec{k}_{\mathrm{gw}}\bot \vec{B}^{0}$). Since the {\sc gw} amplitude depends only on $z-t$, we can interpret the vertical axis either as the propagation direction $z$ at a certain time $t_{0}$ or as the temporal evolution for a ring of particles around the spatial origin $z=z_{0}$ as in a Minkowski diagram. In the absence of the {\sc gw} the world-sheet would be a straight cylinder, but due to the {\sc gw} the circular circumferences are periodically stretched and compressed into ellipses with axes along the $x$ and $y$ axis for this particular $+$ {\sc gw} polarisation. The corresponding behaviour of a magnetic field that was spatially uniform in the absence of the {\sc gw}, is a periodic amplification of the magnetic field when $B_x^0$ is amplified by the {\sc gw} ($B_x^1>0$) and dilution when $B_x^0$ is suppressed ($B_x^1< 0$). The result is a magneto-compressional {\sc msw} mode.

%--
\subsection{Magnetosonic waves in a Poynting flux wind}
When the plasma is tenuous and strongly magnetised such that the magnetic pressure greatly exceeds the gas pressure, the sound velocity is much smaller that the relativistic Alfv{\'e}n velocity which approaches the speed of light  $u_{\mathrm{A}}\simeq c$. In this {\em Poynting flux dominated} limit we can neglect the gas pressure and the {\sc msw} and the {\sc gw} obey almost the same dispersion relation $\omega_{\mathrm{gw}} = \omega_{\mathrm{msw}} = k_{\mathrm{gw}} c = k_{\mathrm{msw}} u_{\mathrm{A}}$ and can interact coherently over the longest interaction scales.
%%%%%%%%%%%%
\begin{figure}
\resizebox{\hsize
}{!}{\includegraphics[angle=-90]{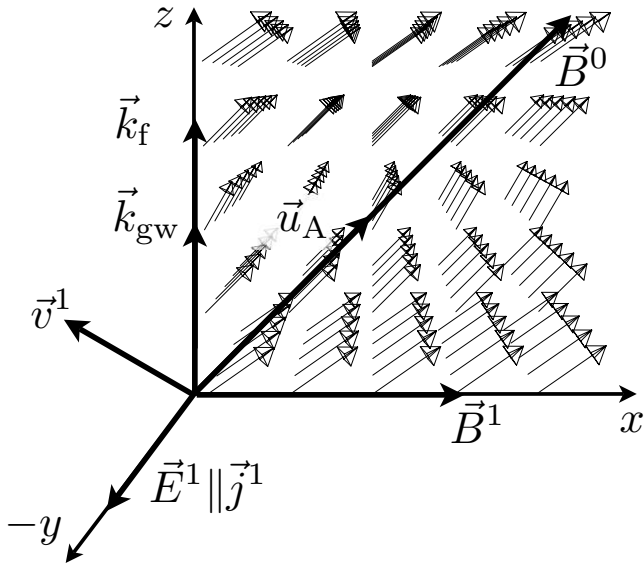}}
\caption{\label{fig::vectorfields}The orientation of the {\sc msw} components and the perturbed magnetic vector field (inset).}
\end{figure}
%%%%%%%%%%%%

The orientation of the different components of the fast {\sc msw} are shown in Fig.~\ref{fig::vectorfields}. The polarisation is transverse, such that the electromagnetic field components and the wave vector form a mutually orthogonal triad: $\vec{E}^{1} \bot \vec{B}^{1}\bot\vec{k}$.
Also shown is a schematic illustration of the perturbed magnetic field for on oblique magneto-acoustic wave propagating in the $z$-direction. The perturbations are exaggerated to emphasise the effect. Since $B_z$ is constant while $B_x$ oscillates, the total magnetic field has an overall wavy pattern.

As was mentioned in the Introduction, the magnetosonic waves excited by a {\sc gw} have frequencies in the kHz regime and cannot escape the plasma directly. However, upon inverse Compton scattering of the {\sc msw} on relativistic particles with a typical Lorentz factor $\gamma$ of a few hundreds, the photon frequency can be boosted by up to a factor $2\gamma^{2}$ larger than that of the incident {\sc msw} wave and into the {\sc lofar} band ($30$--$240$~MHz). 
Close to the source, such frequencies are still well below the local plasma frequency as determined by the pair plasma expelled by the source which effectively behaves as a single millisecond pulsar with a strong magnetic field. The region where scattering into escaping radiation is most likely to occur is far away from the pulsar where the plasma frequency has dropped sufficiently (see also Section~\ref{sec::discussion}).

%%%%%%%%%
\begin{figure}
\resizebox{\hsize}{!}{\includegraphics{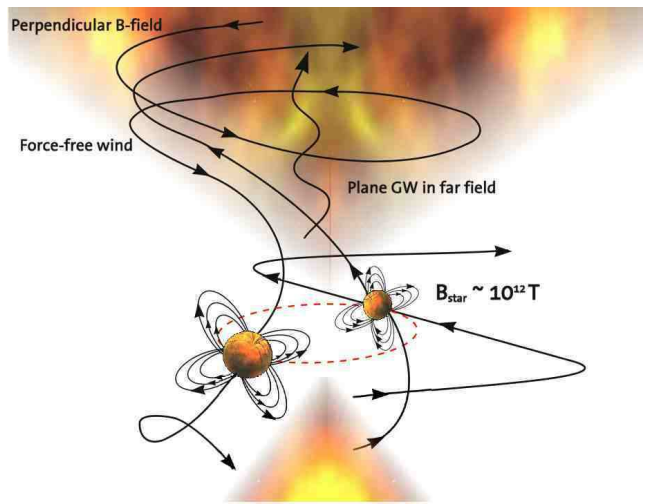}}
\caption{\label{fig::gwbdance}Merging neutron star binary}
\end{figure}
%%%%%%%%%%%%%%%%%%%%%%%%%%%%%%%

In the following sections we will first summarise the covariant description of some fundamental ingredients of relativistic plasma dynamics and then derive the probabilities of scattering a {\sc msw} into the higher frequency normal modes of an intrinsically relativistic plasma. Note that while in the context of this section it was most illustrative to choose the {\sc gw} wave vector along the $z$ axis and the magnetic field at some angle to it, in the remainder of this paper it will be more convenient to do the opposite and choose $\vec{B}^{0}$ along the $z$ axis and the wave vectors in arbitrary directions.

%%%%%%%%%%%%%%%%%%%%%%%%%%%%%%%
\section{Emission}\label{sec::emission}
%%%%%%%%%%%%%%%%%%%%%%%%%%%%%%%

%%%%%%%%%%%%%%%%%%%%%%%%%%%%%%%
\subsection{Wave equation and photon propagator}
%%%%%%%%%%%%%%%%%%%%%%%%%%%%%%%
We will study the emission due to the external current 
generated by a passing {\sc mhd} wave. The inhomogeneous wave equation for the scattered field of arbitrary wave mode $P$ with vector potential $A^{\mu}_{\mathrm{P}}$ is given by:
\bq\label{eq::wave}
\Lambda^{\mu\nu} (k) A_{\mathrm{P}\nu} (k) = -\mu_{0}J_{\mathrm{ext}}^{\mu} (k)\ ,
\eq
where $\Lambda^{\mu\nu}$ is the wave operator for the homogeneous equation which determines the polarisation properties of wave mode $P$ and includes the induced self-consistent field. 
In the {\em weak turbulence} regime one can expand the current excited by a field $A^{\mu}(k)$
in terms of a power series in this field. The first two terms in this expansion define the linear and quadratic non-linear response:
\bqa\nonumber
J^{\mu}_{\mathrm{ind}}(k) &=& \alpha^{\mu}_{\ \nu}(k) A^{\nu}(k) +  \!
\int \frac{d^{4}k_{1}}{(2\pi)^{4}} \frac{d^{4}k_{2}}{(2\pi)^{4}} \delta^{4}(k\!-\!k_{1}\!-\!k_{2})\\
&\times& (2\pi)^{4} \alpha^{\mu}_{\ \nu\alpha} (-k, k_{1}, k_{2}) A^{\nu}(k_{1}) A^{\alpha} (k_{2})\ ,
\eqa
where $k_{i}$ stands for the $4$-vector $(\omega_{i}, \vec{k}_{i})$. 
A plasma theory is different from a vacuum theory in that the linear response of the plasma is included in the wave operator:
\bq\label{eq::waveresponse}
\Lambda^{\mu\nu} (k)= k^{2} g^{\mu\nu} - k^{\mu}k^{\nu} + \mu_{0} \alpha^{\mu\nu} (k)\ ,
\eq
which follows from Maxwell's equations $i k_{\nu}F^{\mu\nu} = \mu_{0}J^{\mu}$
with $F^{\mu\nu} = i k^{\mu} A^{\nu} - i k^{\nu} A^{\mu}$.
The homogenous wave equation (\eqref{eq::wave} with the external current set to zero), and \eqref{eq::waveresponse} for the wave operator then determine the wave eigenmodes of the plasma.

In Section~\ref{sec::current} we will calculate the external current excited by a ({\sc gw} induced) magnetosonic wave propagating through a magnetised relativistic plasma and in Section~\ref{sec::responses} we will derive the response of the plasma to this current. To find the associated scattered wave field, \eqref{eq::wave} has to be inverted:
\bq
A^{\mu}_{\mathrm{P}} (k) = - D^{\mu}_{\ \nu} (k) J_{\mathrm{ext}}^{\nu} (k)\ .
\eq
$D^{\mu\nu}$ is the Green's function for the inhomogeneous wave equation or the {\em photon propagator}. We are interested in the {\em radiation field} excited by the external current so we only consider the irreversible anti-Hermitian part of the photon propagator which is obtained by imposing the causal condition on it and using the Plemelj formula.

The covariant and {\em gauge independent} expressions for the photon propagator and its anti-Hermitian part are\footnote{In \eqref{eq::DA} and throughout this paper, inner products are denoted as $G k = g_{\mu\nu} G^{\mu} k^{\nu}$.}:
\bqa
D^{\mu\nu} (k) &=& \mu_{0}\frac{G_{\alpha} G_{\beta}^{\prime}}{(G k)(G^{\prime} k)} \frac{\lambda^{\mu\alpha\nu\beta}(k)}{\lambda (k)}\ ,\\\label{eq::DA}
D^{\mathrm{A}\mu\nu} (k) &=& -i\pi \mu_{0}\frac{G_{\alpha} G_{\beta}^{\prime}}{(G k)(G^{\prime} k)} \lambda^{\mu\alpha\nu\beta}(k) \delta[\lambda (k)]\ ,
\eqa
where $\lambda^{\mu\alpha\nu\beta}$ and $\lambda (k)$ are the second-order matrix of cofactors and the determinant of $\Lambda^{\mu\nu}$, respectively. $G^{\mu}$ determines the gauge and can be chosen, for instance, as the Lorentz gauge $G_{\mu} = k_{\mu}$, Coulomb gauge $G_{\mu} = [0, \vec{k}]$ or temporal gauge $G_{\mu} = [1, \vec{0}]$.

We will specify to the temporal gauge as it is the most convenient and the only gauge that allows a unique normalisation of the wave polarisation $4$-vectors: one is free to specify $e_{M}^{\mu} (k) e_{M\mu}^{\star} (k)=-1$ with $e_{M}^{\mu}=[0,\vec{e}_{M}]$.
Furthermore, in the temporal gauge the first-order matrix of cofactors and the determinant of the $3$-tensor $\Lambda^{i}_{\ j}$ are given by $\lambda^{i}_{\ j} = \lambda^{0i}_{\ \ 0j}$ and 
$\lambda^{(t)} (k) =\omega^{2} \lambda (k)$, respectively, with  
$\lambda^{i}_{\ l}\Lambda^{l}_{\ j}  = \lambda^{(t)} \delta^{i}_{\ j}$. 
The polarisation vectors can be derived from the spatial matrix of cofactors by
\bq\label{eq::polardef} 
e^{\star}_{M i} (k) e_{M j} (k) = \frac{\lambda^{i}_{\ j}(k)}{\lambda^{i}_{\ i}(k)}\ .
\eq

The photon propagator only has contributions from the poles at the zeros of $\lambda (k_{M})$ corresponding to the dispersion relation $k=k_{M}$ for a specific wave mode. 
To make this more explicit we use the identity:
\bq\label{eq::deltaexpand}
\delta[\lambda (k)] = \sum_{M} \frac{\delta[\omega\!-\!\omega_{M} (\vec{k})] \!+ \!\delta[\omega\!+\!\omega_{M} (-\vec{k})]}{\left| \partial\lambda(k)/\partial\omega \right|}\ ,
\eq
which can be written in terms of the ratio of electric to total energy
\bq\label{eq::RM}
R_{M} (k) = \frac{W^{(E)}_{M}(k)}{W_{M}(k)}= 
\left.\frac{-\omega}{\frac{\partial\lambda (k)}{\partial\omega}}\right|_{\omega= \omega_{M}} \!= 
\left.\frac{\lambda^{i}_{\ i}(k)}{\omega\frac{\partial \lambda^{(t)}(k)}{\partial \omega}}\right|_{\omega= \omega_{M}}\ .
\eq
Here the wave energy $W_{M}(k) = P^{0}_{M}(k) = T^{00}_{M}(k)$ is the temporal component of the $4$-momentum $P^{\mu}_{M}(k)$ and the time-time component of the energy-momentum tensor $T^{\mu\nu}_{M}(k)$.

Combining \eqref{eq::DA}--\eqref{eq::RM}, the radiation field in mode $M$ is found by retaining only the resonant anti-Hermitian part of the photon propagator, given by:
\bqa\nonumber
D^{\mathrm{A}\mu\nu}_{M} (k) &=& -i\pi \mu_{0} \frac{R_{M}(k)}{\omega_{M}} \left\{e^{\mu}_{M} (k) e^{\star\nu}_{M} (k) \delta[\omega - \omega_{M}(\vec{k})]\right. \\\label{eq::propagator}
 &+& \left. e^{\star\mu}_{M} (k) e^{\nu}_{M} (k) \delta[\omega + \omega_{M}(-\vec{k})]\right\}\ ,
\eqa
while $D^{\mathrm{A}\mu\nu} (k) = \sum_{M} D^{\mathrm{A}\mu\nu}_{M} (k)$ and since \eqref{eq::propagator} is derived using the temporal gauge, only the spatial components $(D^\mathrm{A}_{M})^{i}_{\ j}$ are non-zero.

%%%%%%%%%%%%%%%%%%%%%%%%%%%%%%%
\subsection{Emission probability and $4$-momentum radiated}
%%%%%%%%%%%%%%%%%%%%%%%%%%%%%%%
The {\em $4$-momentum} radiated in an arbitrary wave mode due to an arbitrary source, described by an extraneous current, is identified by the work done by this current on the electromagnetic field. Written covariantly and averaged over a truncation or normalisation time $T$ (which we can let go to infinity), using the power theorem and the charge continuity equation $k^{\nu} J_{\nu}=0$, one has:
\bqa\label{eq::probability}
\frac{1}{T} \int d^{4}x\ J_{\mathrm{ext}}^{\nu} (x) F_{\nu}^{\ \mu}(x) &=& \\\nonumber
\frac{1}{T} \int \frac{d^{4}k}{(2\pi)^{4}}\ \Re [-ik^{\mu} J_{\mathrm{ext}}^{\nu} (k) A_{\nu}(k)]
&=& \int \frac{d^{3}\vec{k}}{(2\pi)^{3}}\ Q_{M}^{\mu}(k)\ .
\eqa
$Q_{M}^{\mu}(k) \equiv dP^{\mu}_{M}(k)/dt$ is the rate at which $4$-momentum is radiated into mode $M$. Using \eqref{eq::wave} with \eqref{eq::propagator} and performing the integral over the delta functions we find:
\bqa\nonumber 
Q_{M}^{\mu}(k)&=& \frac{\mu_{0}R_{M}(k)}{T\omega_{M}} k^{\mu}_{M} \left| e^{\star}_{M\nu}(k) J_{\mathrm{ext}}^{\nu} (k_{M})\right|^{2} \\\label{eq::probdef}
&\equiv& k^{\mu}_{M} w_{M} (k)\ ,
\eqa
Equation \eqref{eq::probdef} defines the probability $w_{M} (k)$ per unit time and unit volume of $\vec{k}$-space of spontaneous emission of a wave quantum.

The rate of emission $Q_{M}^{\mu}$ and the ratio of electric to total energy $R_{M}$ can also be written more conveniently in terms of the eigenvalues of the response tensor 
\bq\label{eq::a}
\alpha_{M} (k) = e^{\star}_{M\mu} (k) e_{M\nu} (k) \alpha^{\mu\nu} (k)\ ,
\eq
such that
\bsub
\begin{align}\label{eq::RMII}
\frac{1}{R_{M} (k)} &= \left[2-\frac{\mu_{0}}{\omega}\frac{\partial}{\partial\omega}
\alpha_{M} (k)\right]_{\omega=\omega_{M}}\ ,
\\\label{eq::qm}
Q_{M}^{\mu} (k) &= -  \frac{2i\mu_{0} R_{M} (k) W_{M} (k)}{\omega_{M}^{2}} k_{M}^{\mu} \alpha^{A}_{M} (k_{M})\ ,
\end{align}
\esub
where the purely imaginary anti-Hermitian part of \eqref{eq::a}, $\alpha^{A}_{M} (k_{M})$, measures the effect of dissipation in mode $M$.

%%%%%%%%%%%%%%%%%%%%%%%%%%%%%%%
\section{GW induced current}\label{sec::current}
%%%%%%%%%%%%%%%%%%%%%%%%%%%%%%%

%%%%%%%%%%%%%%%%%%%%%%%%%%%%%%%
\subsection{Zeroth order current}
%%%%%%%%%%%%%%%%%%%%%%%%%%%%%%%
The zeroth-order motion of electrons and positrons in a uniform and static magnetic field is a spiralling orbit around the magnetic field given by the covariant equation of motion:
\bq\label{eq::eom}
\frac{d u^{\mu} (\tau)}{d\tau} = \frac{q}{m} F_{0}^{\mu\nu} u_{\nu} (\tau)\ ,
\eq
where $q$ is the charge of the particle, $m$ the mass and $\tau$ the proper time. One can define a magnetostatic field as having $F^{\mu\nu}F_{\mu\nu} >0$ in any frame in which case it is possible to choose a particular frame where $\vec{E} = 0$ and $\vec{B} \neq 0$. We will orient the axes such that the magnetic field lies along the $z$-axis and define $B= \half\sqrt{F_{0}^{\mu\nu}F_{0\mu\nu}}$. 
If we specify the initial condition for the $4$-velocity as $u_{0}^{\mu} = \gamma (1, v_{\bot}, 0, v_{\|})$ at $\tau =0$ we can integrate Eq.~\ref{eq::eom} twice to find the orbit\footnote{Note that $u_{0}^{\mu}$ denotes an initial condition, whereas $u^{(0)\mu}$ means the unperturbed $4$-velocity and similarly for $X$.}:
\begin{subequations}\label{eq::orbit0}
\bqa
u^{(0)\mu} (\tau) &=& \dot{t}^{\mu\nu} (\tau) u_{0\nu}\ , \\\label{eq:zerothorbit}
X^{(0)\mu} (\tau) &=& x_{0}^{\mu} + t^{\mu\nu} (\tau) u_{0\nu} \ ,
\eqa
\end{subequations}
where $t^{\mu\nu}(\tau)$ and its Fourier transform $\tau^{\mu\nu} (\omega)$ are given by:
\bqa\nonumber
t^{\mu\nu} (\tau) &=& \frac{1}{\Omega_{c}}\left(
\ba{cccc}
\Omega_{c}\tau & 0 & 0 & 0\\
0 & - \sin\Omega_{c}\tau & \eta \cos\Omega_{c}\tau & 0\\
0 & - \eta\cos\Omega_{c}\tau & - \sin\Omega_{c}\tau & 0\\
0&0&0&-\Omega_{c}\tau
\ea
\right)\ ,\\ 
\tau^{\mu\nu} (\omega) &=& \left(
\ba{cccc}
1 & 0 & 0 & 0\\
0 & - \frac{\omega^{2}}{\omega^{2} - \Omega_{c}^{2}} & - \frac{i\eta \Omega_{c} \omega}{\omega^{2} - \Omega_{c}^{2}} & 0\\
0 & \frac{i\eta \Omega_{c} \omega}{\omega^{2} - \Omega_{c}^{2}} &-\frac{\omega^{2}}{\omega^{2} - \Omega_{c}^{2}} & 0\\
0&0&0&-1
\ea
\right)\ .
\eqa 
$\gamma$, $v_{\bot}$, $v_{\|}$ and the components of  the $4$-momentum $p^{\mu} = m u^{\mu}$, $p_{\bot} = \gamma m v_{\bot}$ and $p_{\|} = \gamma m v_{\|}$ are all constants of the motion. Also, we have defined $\eta \equiv q/|q|$ and the (non-relativistic) gyro-frequency $\Omega_{c} = \left|q B/m\right|$.

Note that in the limit $\Omega_{c}\rightarrow 0$, $\tau^{\mu\nu} (\omega)$ reduces to the Minkowskian metric tensor $g^{\mu\nu} = \eta^{\mu\nu}$ and \eqref{eq::orbit0} gives rectilinear motion.

The {\em zeroth-order} single particle current in the absence of any perturbations is now given in Fourier space by:
\bq
J^{\mu}_{\mathrm{sp}} (k)  = q \int d \tau\ u^{\mu} (\tau) \mathrm{e}^{i k X (\tau)}\ .
\eq

%%%%%%%%%%%%%%%%%%%%%%%%%%%%%%%
\subsection{First-order current}\label{sec::current1}
%%%%%%%%%%%%%%%%%%%%%%%%%%%%%%%
We now study a particle that is orbiting in a static magnetic field when an {\sc mhd} wave with amplitude $A^{\mu}(k)$ passes and perturbs the motion, by including the covariant Lorentz acceleration $S^{\mu}$ in \eqref{eq::eom}: 
\begin{subequations}\label{eq::eom1}
\bqa
\frac{d u^{\mu} (\tau)}{d\tau} &=& \frac{q}{m} F_{0}^{\mu\nu} u_{\nu} (\tau) + S^{\mu} (\tau)\ ,\\\nonumber
S^{\mu} (\tau) &=& 
\frac{i q}{m} \int \frac{d^{4} k_{1}}{(2\pi)^{4}} \mathrm{e}^{-i k_{1} X (\tau)} \\\label{eq::lorentz}
&\times& k_{1} u(\tau) G^{\mu\nu} (k_{1}, u(\tau)) A_{\nu} (k_{1})\ .
\eqa
\end{subequations}
where the useful projection tensor $G^{\mu\nu}$ is defined by:
\bq
G^{\mu\nu} (k, u) = g^{\mu\nu} - \frac{k^{\mu} u ^{\nu}}{k u}\ .
\eq
The orbits can again be obtained by integrating the first-order equation of motion \eqref{eq::eom1}:
\begin{subequations}\label{eq::orbit1}
\begin{align}
u^{(1)\mu} (\tau) &= \int_{0}^{\tau}\! d\tau^{\prime}\ \dot{t}^{\mu\nu} (\tau - \tau^{\prime}) S^{(0)}_{\nu} (\tau^{\prime})\ ,\\
X^{(1)\mu}(\tau) &= \int_{0}^{\tau}\! d\tau^{\prime\prime}\int_{0}^{\tau^{\prime\prime}}\! d\tau^{\prime}\ \dot{t}^{\mu\nu} (\tau^{\prime\prime} \!-\! \tau^{\prime}) S^{(0)}_{\nu} (\tau^{\prime})\ ,
\end{align}
\end{subequations}
where $S^{(0)}_{\nu} (\tau)$ is given by inserting the zeroth-order orbit Eq.~\ref{eq::orbit0} in Eq.~\ref{eq::lorentz}.

The first-order expansion for the current density is:
\bq\label{eq::jsp}
J_{\mathrm{sp}}^{(1)\mu} (k) = 
q\int d\tau\ j^{(1)\mu}(\tau) \mathrm{e}^{i k X^{(0)}(\tau)}\ ,
\eq
with:
\bq\label{eq::tmp}
j^{(1)\mu}(\tau) = u^{(1)\mu}(\tau) + i k X^{(1)}(\tau) u^{(0)\mu}(\tau)\ .
\eq
Inserting \eqref{eq::tmp} in \eqref{eq::jsp}, partially integrating with $u^{(1)\mu} = dX^{(1)}/d\tau$ and using \eqref{eq::orbit1}, results in:
\bqa\nonumber
J_{\mathrm{sp}}^{(1)\mu} (k) &=& i q \int d\tau\ k u^{(0)}(\tau) G^{\alpha\mu}(k, u^{(0)}(\tau) ) X^{(1)}_{\alpha}\mathrm{e}^{i k X^{(0)}}
\\\label{eq::singleparticle}
 &=& -\frac{q^{2}}{m} \int\!d\tau\int_{0}^{\tau}\!d {\tau^{\prime\prime}}\int_{0}^{\tau^{\prime\prime}}\!d\tau^{\prime}
\int\frac{d^{4} k_{1}}{(2\pi)^{4}}\\\nonumber
&\times& \mathrm{e}^{i\left[k X^{0} (\tau)-  k_{1} X^{0} (\tau^{\prime})\right]}\dot{t}_{\alpha\beta}(\tau^{\prime\prime} - \tau^{\prime})A_{\nu} (k_{1}) \\\nonumber
&\times& k u^{0}(\tau) G^{\alpha\mu} (k, u^{0}(\tau)) 
k_{1} u^{0}(\tau^{\prime}) G^{\beta\nu} (k_{1}, u^{0}(\tau^{\prime}))\ .
\eqa
Eq.~\eqref{eq::singleparticle} is the current associated with a single particle. 
To calculate the response tensor for the magnetised plasma one can use either the {\em forward-scattering} method were all the perturbations are included in the orbits of the particles and an average is made over the initial conditions,
or the {\em Vlasov method} in which the perturbations are included in the distribution function. In an {\em unmagnetised plasma}, the simplest method is to calculate the 
response tensor for a cold plasma fluid and generalise this result to an arbitrary distribution function by using Lorentz transformations. This method is generally not suitable for a magnetised plasma since there is no inertial frame associated with the gyrating particles. An exception is the case where the particles move strictly along the magnetic field. Since we want to study the effect of a {\sc mhd} wave propagating through the force-free plasma wind relatively close to the source where the magnetic field is very strong and any transverse momentum is quickly synchrotron radiated, a {\em strictly parallel particle distribution} should be a reasonable assumption. 

One advantage of the small gyro-radius approximation is that while \eqref{eq::singleparticle} generally has to be expanded in Bessel functions ($J_{s}(k_{\bot}R)$ with $R$ the gyro-radius) to carry out the integrals, in the strictly parallel case the arguments of all the Bessel functions vanish for $v_{\bot}\propto R =0 $ which leads to $J_{s} (0) =0$ for all integers $s\neq 0$ and $J_{0} (0) =1$, and
\bqa\nonumber
J_{\mathrm{sp}}^{(1)\mu} (k) &=&-\frac{q^{2}}{m} \int\frac{d^{4} k_{1}}{(2\pi)^{4}} \mathrm{e}^{-i(k -k_{1})x_{0}} 
 2\pi \delta\left[(k_{\|}-k_{1\|})u\right] \\\label{eq::thecurrent} &\times&
 G^{\alpha\mu} (k, u) \tau_{\alpha\beta}(k_{\|} u)
 G^{\beta\nu} (k_{1}, u)A_{\mathrm{M}\nu} (k_{1})\ , 
\eqa
where $A_{M}^{\mu}$ denotes the incident fast magneto-acoustic {\sc mhd} wave and since $u$ is strictly along the magnetic field we implicitly have $k u = (k u)_{\|}= \gamma (\omega - k_{\|}u_{\|})$, $v_{\bot}=0$.

%%%%%%%%%%%%%%%%%%%%%%%%%%%%%%%
\section{Linear and non-linear response}\label{sec::responses}
%%%%%%%%%%%%%%%%%%%%%%%%%%%%%%%
To find the linear response tensor for the magnetised plasma, we use the forward-scattering method and average the first-order single-particle current \eqref{eq::singleparticle} over the distribution of particles $F(x_{0}, p_{0})$.  The distribution function is defined covariantly in terms of the number $d{\cal N}$ of world lines --one per particle-- in an $8$-dimensional phase space threading a $7$-dimensional surface element $d^{4}x_{0}d^{4}p_{0}/d\tau$ as determined by: $d{\cal N}d\tau = F(x_{0}, p_{0}) d^{4}x_{0}d^{4}p_{0}$. With this identification one can replace the integral over $d\tau$ in \eqref{eq::singleparticle} with an integral over $d{\cal N}d\tau$ --that is over $d^{4}x_{0}d^{4}p_{0}$ weighed by $F(x_{0}, p_{0})$-- to average over the ensemble of particles. 

When the particle distribution is uniform in space and time, $F(p_{0})$ does not depend on the initial conditions $x_{0}$ and the current \eqref{eq::singleparticle} only depends on $x_{0}$ through the unperturbed orbit \eqref{eq:zerothorbit}. Explicitly, this results in a factor $\exp[i(k-k_{1}) x_{0}]$ as in \eqref{eq::thecurrent} and the integral over $x_{0}$ gives a delta function $(2\pi)^{4}\delta^{4}(k-k_{1})$ over which the $k_{1}$ integral is performed.

An alternative and easier procedure is to use \eqref{eq::thecurrent} which is valid in the small gyro-radius approximation, realise that $2\pi \delta\left[(k_{\|}-k_{1\|})u\right]=\int d\tau \exp\left[i (k_{\|}-k_{1\|})u\tau\right]$ and replace this integral with the integral over phase space. The result is:

\bqa\nonumber
\alpha^{\mu\nu} (k) &=& -\frac{q^{2}}{m}\int d^{4}p\ F(p)G^{\alpha\mu} (k, u) \tau_{\alpha\beta} (k u) G^{\beta\nu} (k,u)\\\label{eq::alphak}
&=&-\frac{q^{2}}{m}\int\frac{d p_{\|}}{\gamma_{\|}} g(p_{\|}) \alpha^{\mu\nu} (k, u)\ ,
 \eqa 
where the last equality defines $\alpha^{\mu\nu} (k, u)$ and the distribution $F(p)$ is chosen parallel to the magnetic field, so the spatial part can be written as: $f(\vec{p}) = \delta^{2}(\vec{p}_{\bot}) g(p_{\|})$. More specifically, a relativistic thermal distribution strictly parallel to the magnetic field is given by the one-dimensional J{\"u}ttner distribution:
\bsub\label{eq::juttner}
\bqa
F(p) &=& 2\pi m \delta(p^{2} - m^{2})  \delta^{2}(\vec{p}_{\bot}) g(p \tilde{u})\ , \\\label{eq::juttner1D}
g(\gamma) &=&  \frac{n \mathrm{e}^{-\frac{m \gamma}{T}}}{2\pi m K_{1}(m/T)}\ ,
\eqa
\esub
where $K_{\nu}$ are the modified Bessel or Macdonald functions, $T$ is the temperature in energy units, $\tilde{u}$ is the $4$-velocity of the rest frame in which $\tilde{u} = [1, \vec{0}]$ such that $p \tilde{u} = m\gamma$.

The J{\"u}ttner distribution \eqref{eq::juttner} is normalised by the (non invariant) number density $n$ evaluated in the rest frame:
\begin{subequations}
\bq\nonumber
n = \int d^{4}p\ u^{0} F(p) 
=\int_{-\infty}^{\infty}dp_{\|}\ g(\gamma) \ ,
\eq
rather than the invariant proper number density
\bq 
n_{\mathrm{pr}} = \int d^{4}p\ F(p) = \int_{-\infty}^{\infty}dp_{\|}\ \frac{g(\gamma)}{\gamma}\ ,
\eq
\end{subequations}
which does not correspond to the actual number density in any frame (when the plasma is not cold).

In this special case of a strictly parallel distribution \eqref{eq::alphak} can also be obtained from a fluid description as in the unmagnetised case, as was mentioned in the previous section. The covariant linear response tensor for a cold magnetised plasma-fluid with rest frame $4$-velocity $\tilde{u}$ and proper number density $n_{\mathrm{pr}}$ is:
\bq\label{eq::cold}
\alpha^{\mu\nu}_{\mathrm{cold}} (k) = -\frac{q^{2} n_{\mathrm{pr}}}{m} G^{\alpha\mu} (k, \tilde{u}) \tau_{\alpha\beta} (k \tilde{u}) G^{\beta\nu} (k,\tilde{u})\ ,
\eq
and the response tensor \eqref{eq::alphak} for an arbitrary but strictly parallel distribution is found by interpreting $n_{\mathrm{pr}}$ as the proper number density $d^{4}p F(p)$, which implies implicitly that for each particle with $4$-velocity $u$ the response is calculated in its rest frame using \eqref{eq::cold} and subsequently Lorentz boosted {\em along the magnetic field} to the frame of interest and then the contributions of all particles are summed by integrating over $d^{4}p$. 

The cold plasma response \eqref{eq::cold} can be recovered from \eqref{eq::alphak} in the limit $F(p) = n \delta^{4}(p u - m \tilde{u})$ where $n=n_{\mathrm{pr}}$. 

%%%%%%%%%%%%%%%%%%%%%%%%%%%%%%%
\subsection{Linear response of cold magnetised plasma}
%%%%%%%%%%%%%%%%%%%%%%%%%%%%%%%
The explicit spatial components of \eqref{eq::cold} will be useful later on. In terms of the total plasma frequency $\omega_{p}^{2} = \omega_{\mathrm{p}+}^{2} + \omega_{\mathrm{p}-}^{2}$, the positron plasma frequency $\omega_{\mathrm{p}+}^{2} = e^{2} n_{+}/(\epsilon_{0}m)$ (and similarly for the electron plasma frequency) and the average charge number $\xi = (n_{+} - n_{-})/(n_{+} + n_{-})$ measuring the excess of positrons (with density $n_{+}$) to electrons, one has \citep{melrose}:
\bq\label{eq::coldalpha}
\alpha_{\mathrm{cold}}^{ij} = -\frac{\omega_{\mathrm{p}}^{2}}{\mu_{0}} \left(
\ba{ccc}
\frac{\omega^{2}}{\omega^{2} - \Omega_{c}^{2}} & 
\frac{i \xi \omega\Omega_{c}}{\omega^{2} - \Omega_{c}^{2}} & 0 \\
 -\frac{i \xi \omega\Omega_{c}}{\omega^{2} - \Omega_{c}^{2}}& \frac{\omega^{2}}{\omega^{2} - \Omega_{c}^{2}} & 0\\
 0 & 0 & 1
\ea
\right)\ .
\eq
We are interested in a pure pair plasma with $\xi=0$ in which case the linear cold plasma response tensor \eqref{eq::coldalpha} becomes diagonal and the contributions of the electrons and positrons add in the (linear) Thomson scattering process, in contrast to for instance an electron gas which has $\xi=-1$.

More generally, the plasma might have some charge excess, for instance related to the Goldreich-Julian charge density $n_{\mathrm{GJ}}$ such that $\xi \simeq n_{\mathrm{GJ}}/n_{\mathrm{tot}} = 1/M$. $M$ is the {\em multiplicity} of secondary particles created by a primary particle that emits curvature radiation while it flows out along an open field line anchored to the polar cap. The curvature photons then  produce a cascade of pair creation \citep{beskin}. The multiplicity is largely unknown but varies from $10^{2}$--$10^{6}$. The gyrotropic terms only contribute quadratically in the dispersion relation for the plasma, i.e.~of order $1/M^{2}$, which is the motivation to neglect those terms in \citet{gedalin98}, Section~V A. For very low-frequency waves and low multiplicity the gyrotropic terms become important when $ 1/M \simeq \xi  \sim \omega/\Omega_{c}$ as can be seen by comparing the components of \eqref{eq::coldalpha}. Although this might lead to interesting behaviour, we will assume for simplicity that the plasma is sufficiently neutral that $\xi$ can be neglected.

%%%%%%%%%%%%%%%%%%%%%%%%%%%%%%%
\subsection{Linear response of relativistic magnetised pair plasma}
%%%%%%%%%%%%%%%%%%%%%%%%%%%%%%%

In evaluating \eqref{eq::alphak} for more complicated response tensors it is convenient to define the average, $\langle V\rangle$, of any variable $V$ by:
\bq
\langle V\rangle \equiv \frac{1}{n} \int \frac{dp_{\|}}{\gamma} V g(p_{\|})\ .
\eq
The frame in which the bulk momentum vanishes, $\langle\gamma v\rangle =0$, is defined as the rest frame and the non-gyrotropic components can be written 
in terms of the Alfv{\'e}n speed, $v_{\mathrm{A}}$, a mean square speed, $\Delta v^{2}$,
\bq
v_{\mathrm{A}} = \frac{\Omega_{c}}{\sqrt{\langle\gamma\rangle}\omega_\mathrm{p}}\ , \quad
\Delta v^{2} = \frac{\langle\gamma v^{2}\rangle}{\langle\gamma\rangle}\ ,
\eq
and the {\em relativistic plasma dispersion function} for the phase velocity $z = \omega/k_{\|}$ parallel to the magnetic field:
\bq
W(z) = \left<\frac{1}{\gamma^{3}(z-v)^{2}} \right>=\int_{-\infty}^{\infty}\frac{dp}{v-z}\frac{dg(p)}{dp}\ .
\eq

Since we are interested in scattering of kHz waves to $\sim100$~MHz waves that both have $k u\ll \Omega_{c}$, we can expand in $k u/\Omega_{c}$ and only keep terms up to second order. Under these approximations, the linear response tensor is given in terms of the refractive index $n=k/\omega$ (and $n_{\bot} = n\sin\theta$, $n_{\|}=n\cos\theta$) by:
\begin{align}\label{eq::hotalpha}
\alpha^{ij}_{\mathrm{rel}} = 
\frac{-\omega^{2}}{v_{\mathrm{A}}^{2}\mu_{0}}\left(
\ba{ccc}
1+n^{2}_{\|}\Delta v^{2} & 0 & - n_{\bot} n_{\|}\Delta v^{2} \\
 0 & 1+n^{2}_{\|}\Delta v^{2} & 0\\
 -n_{\bot} n_{\|}\Delta v^{2}& 0 & n^{2}_{\bot}\Delta v^{2}- f(n)
\ea
\right)\ ,
 \end{align}
with $f(n) = \frac{\omega_{p}^{2}}{\omega^{2}} v_{\mathrm{A}}^{2} z^{2} W(z)$.

Note that for $\xi = 0$, $\Delta v^{2}\!\downarrow\! 0$, $\gamma = \langle\gamma\rangle\downarrow \!1$, $v_{\mathrm{A}}^{2}\!\rightarrow \!\Omega_{c}/\omega_{\mathrm{p}}$ and consequently $W(z)\rightarrow z^{-2}$, \eqref{eq::hotalpha} reduces to the low-frequency limit of the cold plasma response tensor \eqref{eq::coldalpha} with $\xi =0$.

In the remainder of this paper, we will investigate whether the fast {\sc msw} eigenmode of \eqref{eq::coldalpha} --induced by the {\sc gw}-- can scatter into an eigenmode of \eqref{eq::hotalpha} when it propagates through a relativistic plasma region.

%%%%%%%%%%%%%%%%%%%%%%%%%%%%%%%
\subsection{Non-linear response tensor}
%%%%%%%%%%%%%%%%%%%%%%%%%%%%%%%
The quadratic response tensor for a relativistic magnetised pair plasma is obtained in the forward-scattering approach by averaging the second order (in the field) current over an arbitrary distribution function \citep{melrosebook}. We take the small gyro-radius limit to find:
\begin{align}\label{eq::hotnonlin}
&\alpha^{\mu\nu\rho} (k_{0}, k_{1}, k_{2}) = \sum_{\pm}\frac{\pm e^{3}}{4 m^{2}} \int d^{4} p\ F_{\pm}(p) \times \\\nonumber
&\qquad \times G^{\star\alpha\mu} (k_{0}, u) 
G^{\star\beta\nu}(k_{1}, u)
G^{\star\gamma\rho}(k_{2}, u) \times \\\nonumber
 &\left\{
\frac{k_{1}^{\sigma}}{k_{0\|}u} \tau_{\sigma\alpha} (k_{0\|}u) \tau_{\beta\gamma} (k_{2\|} u) \right. +
\frac{k_{2}^{\sigma}}{k_{0\|}u} \tau_{\sigma\alpha} (k_{0\|}u) \tau_{\beta\gamma} (k_{1\|} u) + \\\nonumber
&\frac{k_{0}^{\sigma}}{k_{1\|}u} \tau_{\sigma\beta} (k_{1\|}u) \tau_{\alpha\gamma} (k_{2\|} u) +
\frac{k_{0}^{\sigma}}{k_{2\|}u} \tau_{\sigma\beta} (k_{2\|}u) \tau_{\alpha\gamma} (k_{1\|} u) + \\\nonumber
&\frac{k_{1}^{\sigma}}{k_{2\|}u} \tau_{\sigma\gamma} (k_{2\|}u) \tau_{\beta\alpha} (k_{0\|} u) +\left.
\frac{k_{1}^{\sigma}}{k_{1\|}u} \tau_{\sigma\beta} (k_{1\|}u) \tau_{\gamma\alpha} (k_{0\|} u)
\right\}\ .
\end{align}
The cold plasma limit can be derived from \eqref{eq::hotnonlin} by choosing $F_{\pm}(p) = n_{\pm} \delta^{4}(p-m\tilde{u})$ (where again $\tilde{u} = [1,\vec{0}]$ in the plasma rest frame). The only components in \eqref{eq::hotnonlin}
that depend on the sign of the charge are the gyrotropic terms which are proportional to $\eta = q/|q|$. 
In a pure pair plasma with equal distributions for the electrons and positrons the non-gyrotropic terms are therefore the only ones that do not vanish in the sum over the species (see also the next Section~\ref{sec::scattering}). 

%%%%%%%%%%%%%%%%%%%%%%%%%%%%%%%
\section{Scattering}\label{sec::scattering}
%%%%%%%%%%%%%%%%%%%%%%%%%%%%%%%
As mentioned previously, the emission probability \eqref{eq::probability} is very general and applies to any emission process due to an arbitrary source expressed as an external current density. The specific probability of scattering of a specific mode $M$, say a {\sc mhd} wave, on an electron into a different mode $P$ can now be calculated by inserting the extraneous single particle current \eqref{eq::singleparticle} into \eqref{eq::probability} and interpreting the resulting emission as the scattered wave:

\begin{subequations}\label{eq::wscat}
\begin{align}\label{eq::scatterprob}
w_{MP} (k, k_{1}, u) &= \frac{3\sigma_{\mathrm{T}}(4\pi)^{2}}{\gamma}\frac{R_{M} (k) R_{P} (k_{1})}{\omega_{M} (k) \omega_{P} (k_{1})} \\\nonumber
&\times |a_{MP}(k, k_{1}, u)|^{2} \delta\left[(k_{M\|} - k_{1P\|}) u\right]\ ,\\\label{eq::nonlin} 
a_{MP}(k, k_{1}, u) &= e^{\star}_{M\mu} (k) e_{P\nu} (k_{1})
 \left\{\right.\\\nonumber
  &G^{\alpha\mu} (k_{M}, u) \tau_{\alpha\beta}(k_{M\|} u)
 G^{\star\beta\nu} (k_{1P}, u)\\\nonumber
 &-\left.
\frac{2m}{q} \alpha^{\mu\nu\rho} (k_{M}, k_{1P},k_{M}- k_{1P}) u_{\rho} \right\}\ ,
\end{align}
\end{subequations}
where we have used $[\delta((k\! -\!k_{1})u)]^{2}=T \delta((k\! -\!k_{1})u)/(2\pi\gamma)$. 
We have included the non-linear scattering probability $\alpha^{\mu\nu\rho} (k_{M}, k_{1P},k_{M}- k_{1P})$ due to the scattering of the {\sc mhd} wave on the self-consistent field associated with the zeroth-order current related to the beat between the {\sc mhd} wave and the scattered wave.  As can be seen from \eqref{eq::wscat} the total scattering probability is proportional to $\sigma_{\mathrm{T}}\propto q^{4}/m^{2}$.  The non-linear scattering probability has an additional (small) factor $2m/q$ as can be seen from \eqref{eq::nonlin} and consequently the gyrotropic terms are the {\em only} terms that {\em do not} vanish when the contributions of the two species are added for a pure pair plasma.
The components of the quadratic response, however, are proportional to $ (\omega/\Omega_{c})^{4}$ and are negligible in our long-wavelength approximation (see also \cite{melrose97, gedalin83, mikhailovskii80}).

%%%%%%%%%%%%%%%%%%%%%%%%%%%%%%%
\subsection{Polarization vectors}
%%%%%%%%%%%%%%%%%%%%%%%%%%%%%%%
The polarisation properties of the incident {\sc mhd} wave and the scattered modes are determined by the eigenvectors of the wave operator $\Lambda^{\mu\nu} (k) $ as defined by \eqref{eq::waveresponse} and \eqref{eq::polardef} with the linear response tensors derived in \eqref{eq::coldalpha} and \eqref{eq::hotalpha}.
When the $z$-axis is chosen along the magnetic field and the wave vectors in the $x-z$ plane, then the spatial parts of both the wave operator for the cold plasma and that for the relativistic an-isotropic plasma are of the form:
\bq
\Lambda_{ij} (k)=
\left(
\ba{ccc}\label{eq::wavy}
\Lambda_{11} & 0 & \Lambda_{13}\\
0 & \Lambda_{22} & 0\\
\Lambda_{13} & 0 & \Lambda_{33}
\ea
\right)\ .
\eq
So the dispersion relations for the different wave modes, used in \eqref{eq::deltaexpand}, follow from:
\bq
|\Lambda_{ij} (k)| = \Lambda_{22}(\Lambda_{11}\Lambda_{33} - \Lambda_{13}^{2}) =0\ ,
\eq
and the polarisation vectors are given by:
\begin{subequations}\label{eq::allpols}
\bqa\label{eq::tpol}
\vec{e}_\mathrm{t} &=& \{0,1,0\}\ ,\\
\vec{e}_{-} &=& \frac{1}{\sqrt{
(\Lambda_{13})^{2}+
(\Lambda_{11})^{2}
}}
\{- {\Lambda_{13}}, 0, {\Lambda_{11}}\}\ ,\\
\vec{e}_{+} &=& \frac{1}{\sqrt{
(\Lambda_{13})^{2}+
(\Lambda_{11})^{2}
}}
\{{\Lambda_{11}}, 0, {\Lambda_{13}}\}\ .
\eqa
\end{subequations}
Clearly, $\vec{e}_\mathrm{t}$ always corresponds to a mode where the electric field is purely transverse to the magnetic field with dispersion relation $\Lambda_{22}=0$, such as the magneto-acoustic waves excited by a gravitational wave. 

The modes $\vec{e}_{\pm}$ that have an electric field component along the magnetic field correspond to the eigenfrequency solutions of 
\bq
(\Lambda_{11} - \Lambda_{33}) \mp \sqrt{(2\Lambda_{13})^{2}+(\Lambda_{11} - \Lambda_{33})^{2}}=0\ .
\eq

%------------------------------------------
\begin{figure}
\resizebox{\hsize}{!}{\includegraphics{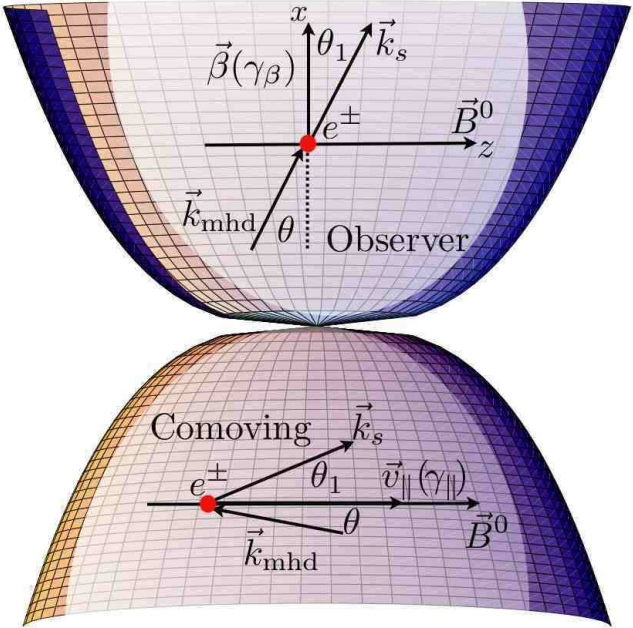}}
\caption{\label{fig::scattering}Scattering geometry in observer (top) and comoving (bottom) frames.}
\end{figure}
%------------------------------------------
%%%%%%%%%%%%%%%%%%%%%%%%%%%%%%%
\subsection{Response to an incident magneto-acoustic wave}
%%%%%%%%%%%%%%%%%%%%%%%%%%%%%%%
To evaluate the scattering probability \eqref{eq::wscat} explicitly, it is most convenient to oriented our coordinate axes such that the background magnetic field is aligned with the $z$-axis and the wave vector of the {\em scattered} wave ($\vec{k}_{P}$) lies in the $x$-$z$ plane. The direction of the {\em incident} magnetosonic wave is then completely arbitrary, but we know from \eqref{eq::wavy} and \eqref{eq::tpol} that it is polarised perpendicular to both the magnetic field and the wave vector. Hence, we have
\begin{subequations}
\bqa
k_{M} &=& \{\omega, k \sin\theta\cos\psi, k \sin\theta\sin\psi, k \cos\theta\}\ ,\\\label{eq::mhdpola}
e_{M} &=& \{0, -\sin\psi,  \cos\psi,0\}\ ,\\
k_{P} &=& \{\omega_{1}, k_{1}\sin\theta_{1}, 0, k_{1}\cos\theta_{1}\}\ ,
\eqa
\end{subequations}
where $\theta$ ($\theta_{1}$) is angle of the incident (scattered) wave vector with respect to the magnetic field and $\psi$ ($\psi_{1}$) the angle with the $x$-axis.
 
Similarly, the polarisation of the scattered modes is found from the eigenvectors of the wave operator \eqref{eq::waveresponse} with the relativistic linear response tensor given by \eqref{eq::hotalpha} using \eqref{eq::allpols}:
\begin{subequations}\label{eq::polas}
\bqa\label{eq::ept}
\vec{e}_{t}(k_{1}) &=& \{0,1,0\}\ ,\\\label{eq::pola1}
\vec{e}_{-}(k_{1}) &=& \frac{1}{\sqrt{1+\Xi^{2}}} \{-\Xi, 0, 1\}\ ,\\\label{eq::pola2}
\vec{e}_{+} (k_{1}) &=& \frac{1}{\sqrt{1+\Xi^{2}}} \{1, 0, \Xi\}\ ,
\eqa
\end{subequations}
with
\bq
\Xi = \frac{k^{2}_{1} u_{A}^{2} \sin\theta_{1}\cos\theta_{1}}{\omega_{1}^{2}-k_{1}^{2} u_{A}^{2} \cos^{2}\theta_{1}}\ .
\eq

With this choice of geometry, the explicit expression for the linear Ôscattering responseÕ summed over the two species such that the gyrotropic terms vanish, is:
\bqa\nonumber
a^{\mu\nu}(k, k_{1}, u)&=&
G^{\alpha\mu} (k_{M}, u) \tau_{\alpha\beta}(k_{M\|} u)
 G^{\star\beta\nu} (k_{1P}, u)\ ,
 \eqa
 with the spatial components written explicitly as:
 \bqa\label{eq::scatterampl}
a_{ij}(k, k_{1}, u) =  - \frac{(k u)^{2}}{(k u)^{2} - \Omega_{c}^{2}} \times &&\\\nonumber
\left(
 \ba{ccc}
1 & 0 & - \frac{\gamma \beta k_{1} \sin\theta_{1}}{k u} \\
 0 &  1 & 0\\
 \frac{\gamma \beta k \sin\theta\cos\psi}{k u} & \frac{\gamma \beta k\sin\theta\sin\psi}{k u} & \Psi
 \ea
 \right)\ ,&&
\eqa
and
$
\Psi = \frac{1}{\gamma^{2}}\frac{\omega}{k u}\frac{\omega_{1}}{k_{1} u} \frac{(k u)^{2} - \Omega_{c}^{2}}{(k u)^{2}} + (1-\gamma^{2})\frac{k\sin\theta\cos\psi}{k u}\frac{k_{1}\sin\theta_{1}}{k_{1} u}\ .
$

%%%%%%%%%%%%%%%%%%%%%%%%%%%%%%%
\subsection{Probability of transverse-transverse scattering}\label{sec::tt}
%%%%%%%%%%%%%%%%%%%%%%%%%%%%%%%
The probability that the incident (transverse) magneto-acoustic wave scatters into a transversely polarised electromagnetic wave, as in \eqref{eq::ept}, is proportional to (the square of):
\bqa\label{eq::suppress}
a_{Mt}(k, k_{1}, u) &=& e^{\star}_{M\mu} (k) e_{t\nu} (k_{1}) a^{\mu\nu}(k, k_{1}, u)\\\nonumber &=&
- \frac{(k u)^{2}}{(k u)^{2} - \Omega_{c}^{2}} = \frac{1}{\frac{\Omega_{c}^{2}}{\gamma^{2}(\omega - k \beta\cos\theta)^{2}}-1}\ .
\eqa
In the two limits of nearly parallel and perpendicular propagation and assuming that $\gamma\omega\ll \Omega_{c}$, this is approximately $\sim\left(\omega/\gamma\Omega_{c}\right)^{2}$ and $\sim\left(\gamma\omega/\Omega_{c}\right)^{2}$, respectively. 

Furthermore, it can be easily seen from the linear response tensors in \eqref{eq::coldalpha} and \eqref{eq::hotalpha}, with the polarisation vectors of the corresponding transverse eigenmodes given in \eqref{eq::mhdpola} and \eqref{eq::ept}, respectively, that the amplitudes
\bq
a_{M,t}(k) = e^{\star}_{M,t\mu}(k) e_{M, t\nu}(k) \alpha^{\mu\nu}(k_{M,t})\left.\right|_{\omega=\omega_{M,t}}\ ,
\eq 
are independent of $\omega$ for the transverse magneto-acoustic and electromagnetic wave modes. Hence, the ratio of electric to total energy given by \eqref{eq::RMII} is $R_{M} =  R_{t} = \half$ just like for transverse modes in an isotropic non-dispersive plasma.  

Compared to the probability of scattering of a fast magneto-acoustic mode to a transverse electromagnetic mode in unmagnetised vacuum, this probability 
is suppressed in a strongly magnetised plasma by the factor in \eqref{eq::suppress} squared. 

%%%%%%%%%%%%%%%%%%%%%%%%%%%%%%%
\subsection{Probability of transverse-LO scattering}
%%%%%%%%%%%%%%%%%%%%%%%%%%%%%%%
The probability that the transverse magneto-acoustic mode is scattered into a quasi-transverse Langmuir-ordinary mode is obtained by contracting \eqref{eq::scatterampl} with the polarisation vectors \eqref{eq::mhdpola} and either \eqref{eq::pola1} or \eqref{eq::pola2}.

The only component of \eqref{eq::scatterampl}, however, that is of lower order in $1/\Omega_{c}$ is the purely longitudinal $zz$ component, which has a term independent of $\Omega_{c}$. This component drops out, though, as soon as \eqref{eq::scatterampl} is contracted with the polarisation vector of the purely transverse incident magnetosonic wave. We find that the probability for scattering into a {\sc lo} mode is equal or less than \eqref{eq::suppress} depending on the relevant angles, the maximum being when the incident wave is perpendicular to the magnetic wave and the electric field of the scattered wave lies in the plane spanned by the electric field of the magneto-acoustic wave and the magnetic field.

\subsection{Emitted power}
In the previous section we derived the probability \eqref{eq::wscat} that one quantum in the {\sc msw} mode scatters into one quantum in another mode. To calculate the actual {\em rate} at which scattered waves are emitted we have to average \eqref{eq::wscat} over the distribution of particles. One subtlety is that the scattering probability per unit time is clearly a frame-dependent quantity. We should therefore integrate either the scattering probability $\gamma w_{MP} (k, k_{1}, u)$ per unit {\em proper time} along the particles motion, multiplied by the {\em proper number density} $d^{4} p F(p)$ or equivalently the probability per unit time over the physical number density $\gamma d^{4} p F(p)$,
similar to the averaging of the response tensor over the distribution function performed in \eqref{eq::alphak}.

Appealing to the semi-classical formalism of detailed balance, we find a kinetic equation for the increase of wave energy in the scattered modes:
\bqa\nonumber
\frac{D W_{P}(k_{1})}{D t} &=&  \int \frac{d^{3}\vec{k}}{(2\pi)^{3}} \int d p_{\|} w_{MP} (k, k_{1}, u) g(p_{\|})\\\label{eq::energyevol}
&\times& \left( \frac{\omega_{1P}}{\omega_{M}} W_{M}(k)-W_{P}(k_{1})\right)\ ,
\eqa
where we only consider spontaneous scattering of $M\rightarrow P$ and $P\rightarrow M$ and neglect induced scattering\footnote{$D/Dt$ is a time-derivative operator that allows for a possible slow variation of the medium itself as $D/Dt = (\partial \omega_{M}/\partial k_{\mu})\partial_{\mu} + (\partial \omega_{M}/\partial x_{\mu})\partial/\partial k^{\mu}$ as in \citet{melrosebook}.}.

The total power in \eqref{eq::energyevol} is calculated for the scattering on the relativistic distribution of particles along the magnetic field in the frame that is comoving with the average bulk flow carrying the magnetic field. Because the power is Lorentz invariant (when the scattered radiation has front-back symmetry in the comoving frame) it is the same in the observer frame. 

However, because of propagation and angular effects there is a difference between the {\em emitted} power $P_{\mathrm{e}}$ and the {\em received} power $P_{\mathrm{r}}$ in the observer frame \citep{ryblight}. The scattered radiation is {\em emitted} during the time it takes the {\sc gw} and {\sc msw} to cross a scale-height, i.e. $\mathcal{H}/c$, but because this shell propagates at a relativistic velocity $\beta$, the radiation is {\em received} in an interval $\mathcal{H}/(2 \gamma_{\beta}^{2} c)$ with a power 
\bq\label{eq::beamed}
P_{\mathrm{r}}\simeq 2 \gamma_{\beta}^{2} P_{\mathrm{e}}\ .
\eq
These relativistic effects also play an important role in the physics of gamma-ray bursts as reviewed in \citet{zhangmesz}. 
Hence the time-integrated emitted and received energies are the same but (with sufficient time resolution) the observed flux is much higher.

%%%%%%%%%%%%%%%%%%%%%%%%%%%%%%%%%%%%%%
\section{Application to NS-NS merger}\label{sec::discussion}%%
%%%%%%%%%%%%%%%%%%%%%%%%%%%%%%%%%%%%%%
We consider the specific environment of a highly relativistic electron-positron wind surrounding the merger of two neutron stars as illustrated in Fig.~\ref{fig::gwbdance}, and apply the general results for the scattering of low-frequency fast {\sc msw}s into escaping electromagnetic waves as derived in the previous sections from the framework in \citet{gedalin98, melrose99, melrosegedalin99,melrose01, luomelrose, melrose02}.  

We assume that the in the last phase of coalescence, the {\em orbital} motion dominates the spin of the individual neutron stars. The orbital angular frequency $\Omega_{b}$ then determines the radius of the light-cylinder $R_{\mathrm{lc}} = c/\Omega_{b}$ where the magnetic field lines that are anchored on the stellar polar caps break open because co-rotation of the plasma which is frozen onto the field lines would result in an unlimited increase in inertia if the filed lines were closed at this distance.
For a more extensive treatment of how the spinning magnetic dipole of a neutron star sets up a voltage jump and an electric field at the stellar polar caps and extracts charges from the neutron star filling the magnetosphere, see for instance \citet{beskin}. Suffice to say that during the last stage of spiral-in, the double neutron star is assumed to behave essentially as a single millisecond pulsar with a charge separated closed magnetosphere and an ultra-relativistic force-free electron-positron plasma wind. 

An important observation is that the relativistic leptonic wind is already present before the actual merger. It doesn't have to be generated by the merger event, which is a problematic issue in collapsar models for gamma-ray bursts, where inside a collapsing massive star a region has to be evacuated of baryons and an almost mass-less jet has to form and then pierce through the entire star without slowing down much or losing its collimation.

As the neutron stars finally merge, a large fraction of the binding energy of the entire system is released in the form of gravitational waves that propagate through the pre-existing strongly magnetised plasma wind. Since a {\em magnetised} perfect fluid possesses an an-isotropic pressure due to the magnetic stress, the {\sc gw} couples to the plasma and excites magnetohydrodynamic waves. In Paper II we found that {\sc gw}s can excite both Alfv{\'e}n and (slow and fast) magneto-acoustic waves but that coupling to the fast mode is the most efficient because the dispersion relations are nearly the same in a tenuous, relativistic magneto-plasma and coherent interaction is always possible.

Even though we proved that the energy transfer from {\sc gw} to {\sc msw} in the plasma can be substantial, we still have to find an efficient radiation process to release this energy. The {\sc msw} cannot escape the plasma directly, so we will now investigate whether the {\sc msw} are likely to inverse Compton scatter into electromagnetic radiation in the radio regime.

\subsection{{\sc msw} amplitude and energy}\label{sec::interaction}
The excitation of fast magneto-acoustic waves by a {\sc gw} propagating through a uniform magnetic field was discussed in Section~\ref{eq::recap}.
The orientation of the ambient magnetic field is now chosen in the $z$-direction such that the wind flows out in the $x$-direction with average bulk velocity $\beta$ and corresponding Lorentz factor $\gamma_{\beta}$. The wave vector of the gravitational waves, $\vec{k}_{\mathrm{gw}}$, is allowed to have an arbitrary angle $\theta$ with respect to the magnetic field (and the wave vector of the excited {\sc msw} will have the same orientation). In reality, the {\sc gw} are emitted more or less radially and $\vec{k}_{\mathrm{gw}}$ will be almost perpendicular to the toroidal magnetic field in the wind in the observer frame (i.e.~as in the top Fig.~\ref{fig::scattering}). 

In the approximation of a uniform background magnetic field, the amplitude and energy of the excited {\sc msw} are given in the observer frame by  
\bsub\label{eq::gwmswenergystuff}
\bqa\label{eq::gwmsw}
B^{(1)}_{z}(x) &\simeq& \frac{h_{+}|B^{(0)}|}{2\gamma_{\beta}^{2}}\sin\theta\ k_{\mathrm{gw}}x\ \mathrm{e}^{i k_{\mathrm{gw}}(x-c t)}\ ,\\
W^{(B)} &=& \int_{V}d^{3}x\ \frac{|B^{(1)}|^{2}}{2\mu_{0}}\ ,\\
W_{M} &=& T^{00}\simeq W^{(E)} + W^{(B)}\simeq 2 W^{(B)}\ .
\eqa
\esub
This result will be generalised to a non-uniform background in the following sections.

\subsection{Eikonal approximation}\label{sec::eikonal}
We generalised \eqref{eq::gwmsw} to a non-uniform magnetoplasma wind by appealing to the {\em eikonal} or {\em geometric optics} two-length-scale expansion (also known as the {\sc wkb} approximation). When the wavelength of the modes under consideration is small compared to the scale of spatial variation in the background\footnote{We assume a time-independent background, but one can obviously also use the eikonal approximation on a time-dependent background when the period of the waves is small compared to the time scale of variations in the background.}, the waves can still be considered locally planar and monochromatic. 
Furthermore, in the Poynting flux dominated regime where the gas pressure is negligible (Section~\ref{sec::interaction}), the classical Alfv{\'e}n velocity $v_{\mathrm{A}}\propto B/\sqrt{n}$ is constant since $B\simeq B_\phi\propto 1/R$ well outside the light-cylinder and $n\propto B_z \propto 1/R^2$  (which will be derived in (\eqref{eq::radial}). Consequently, the relativistic Alfv{\'e}n phase velocity $1/u_{\mathrm{A}}^{2} = 1 + 1/v_{\mathrm{A}}^{2}$ is also constant and the dispersion relation $\omega \simeq k u_{\mathrm{A}}$ is spatially non-dispersive even in the inhomogenous background.

The important space-dependent background quantities are the ambient magnetic field $B^{(0)}(r)$ and plasma density $n(r)$,  and the corresponding gyro-frequency $\Omega_{c}(r)\propto B^{(0)}(r)$ and plasma frequency $\omega_{\mathrm{pl}}(r)\propto \sqrt{n(r)}$. Note that the gravitational waves $h_{+}(r) \exp{i k_{\mathrm{gw}}(r - t)}$ are also treated in the geometric optics approximation as a field living on a flat space-time background, neglecting the curvature due to the neutron star and the (largely magnetic) energy density in the wind. 
The {\sc gw} amplitude is estimated from the total energy released by the merger in {\sc gw} at a distance from the source where we are in the far field approximation, such that the {\sc gw} can be treated as linear, plane wave space-time perturbations. The {\sc gw} amplitude then falls off as $1/r$ and we can scale it such that $h_{+}(r) = h_{\mathrm{lc}} R_{\mathrm{lc}}/r$.

Finally, the linear growth of the excited {\sc msw} perturbation in \eqref{eq::gwmsw} can be generalised to the inhomogenous jet by considering growth over single {\em scale heights} $\mathcal{H}$ (see \eqref{eq::scaleheight}). Putting all of this together, we find
\bsub\label{eq::brs}
\bqa\label{eq::bra}
\frac{B^{(1)}_{\phi}(r)}{B^{(0)}_{\phi}(r)} &\simeq& \frac{h_\mathrm{lc} k_{\mathrm{gw}}R_\mathrm{lc}}{\gamma_{\beta}^{2}}\mathrm{e}^{i k_{\mathrm{gw}}(r- t)}\\\label{eq::brb}
&=& \Psi_{I} \mathrm{e}^{i k_{\mathrm{gw}}(r-t)}\ ,
\eqa
\esub
where the interaction factor $\Psi_{I}$ is now a constant determined by \eqref{eq::bra} and the dependence of $B^{(1)}_{\phi}$, $B^{(0)}_{\phi}$ on $r$ will be determined in the next sections.

Note that because the {\sc gw} and the {\sc msw} both have essentially the same dispersion relation, and the energy flux, $S_1c B^2_\mathrm{mhd}/(4\pi)$, through each surface $S_1$ (as in Fig.~\ref{fig::cones}) is a constant because of \eqref{eq::brb}, both waves can propagate in phase and interact coherently over a long distance even in the non-uniform background.

\subsection{Magnetic field configuration}
The morphology of the `pulsar' wind in which the {\sc gw}--{\sc msw} interaction and the subsequent inverse Compton scattering take place, is fully determined by the magnetic field geometry since the particles are `frozen' to the field lines. 
We assume that the wind is steady and has a constant bulk velocity $v_x$, the magnetic field is axially symmetric and force-free, and we use the fact that the field is solenoidal. 
For such a field, the {\em toroidal} component follows from the torque free condition as $R B_{\phi} =  \mathrm{const.} $ along each field line \citep{lust}, where $(R, \phi, x)$ denote cylindrical coordinates. Note that because of \eqref{eq::brs}, the discussion in this section is valid for both $B^{0}_{\phi}$ and $B^{1}_{\phi}\propto B^{0}_{\phi}$ and we will omit the superscripts to avoid clutter.

\begin{figure}
\centerline{
\resizebox{\hsize}{!}{\includegraphics{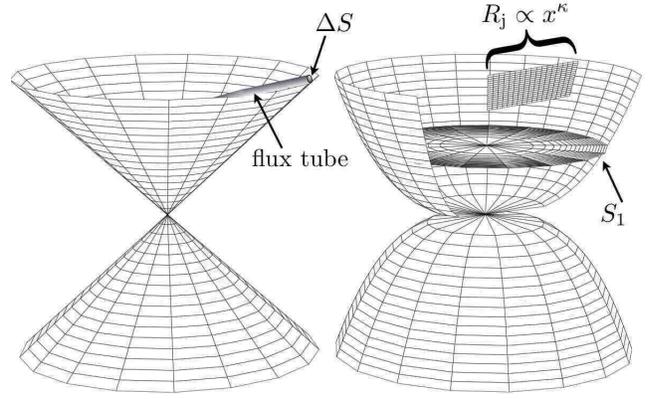}}}
\caption{\label{fig::cones}Wind collimation by the magnetic field for $\kappa = 1$ (left) and $\kappa = \tfrac{1}{3}$ (right).}
\end{figure}

The {\em poloidal} magnetic field $B_{p}$ through any flux tube cross-section $\Delta S$ must be proportional to the particle flux through $\Delta S$ because of flux conservation and the constant flow velocity $v$: $B_{p} \Delta S=  \mathrm{const.}$, $ n v \Delta S=  \mathrm{const.}$, and thus
$
B_{p} \propto n$.
In particular, we will assume that the particle and (poloidal) magnetic flux through an azimuthal cross-section of the {\em entire jet} will approximately be conserved (as illustrated in the right Fig.~\ref{fig::cones}): $B_x S_1 = B_x (\pi R_{\mathrm{j}}^2) \simeq \mathrm{const.}$ 

Summarising, we have
\bq\label{eq::radial}
B_\phi (R) \propto \frac{1}{R}\ , \quad B_x (R_{\mathrm{j}}(x)) \propto n(R_{\mathrm{j}}(x))  \propto\frac{1}{R_{\mathrm{j}}^2}\ ,
\eq
where $R_{\mathrm{j}}(x)$ is the radius of the jet which is related to the opening angle by $\theta_{\mathrm{o}}(x) =2 \arctan R_{\mathrm{j}}(x)/x$.
%------------------------------------------
%------------------------------------------

\subsection{Parametrised wind collimation}
In Papers~I--III we have discussed different degrees of collimation of the wind. We can generalise these results by parametrising the collimation as in Fig.~\ref{fig::cones}, such that the radius $R_{\mathrm{j}}$ of the jet as a function of the vertical distance $x$ is given by $R_{\mathrm{j}} = x^{\kappa}$. From \eqref{eq::radial} then follow the more useful expressions for the vertical dependence 
\bsub\label{eq::parameter}
\begin{align}
B_\phi &\propto x^{-\kappa}\ ,& \quad B_x  &\propto  x^{-2\kappa}\ ,\\\label{eq::vol}
n &\propto x^{-2\kappa}\ ,& V = \int \pi R^{2} d x &\propto x^{2\kappa + 1}\ . 
\end{align}
\esub
We can define a scale-height $\mathcal{H}$ (as in Section~\ref{sec::eikonal}) by the distance over which the toroidal magnetic field falls off to half its strength: 
\bq\label{eq::scaleheight}
\frac{B_{\phi}(x + \mathcal{H})}{B_{\phi}(x)} = \tfrac{1}{2} \quad\Rightarrow\quad
\mathcal{H} \equiv (2^{\tfrac{1}{\kappa}}-1)x\ .
\eq

\subsubsection{$\kappa = 1$}\label{sec::kappa1}
In Paper~I we assumed a wide conical wind with $\kappa = 1$ (see also \cite{kuijpers2001}) as depicted in Fig.~\ref{fig::cones} (left) which has
\bsub
\begin{align}
 B_\phi \propto \frac{1}{r}\ , &\quad B_x \propto \frac{1}{r^2}\ ,\\\label{eq::wkappa}
   W_{M}\propto \frac{1}{r^2}\ , &\quad V  \propto r^{3}\ ,
\end{align}
\esub
 with the radial distance $r = \sqrt{R^2 + x^2} \simeq x$.
The volume integrated energy increases linearly with the interaction length-scale set by the scale-height $\mathcal{H} = r$ from \eqref{eq::scaleheight}. We will only consider the energy deposited in the last scale-height of the force-free wind, as the radiation will escape most easily there.

For a magnetar strength surface magnetic field of $B_{\star} \sim 10^{12}$~T, a millisecond orbit, a Lorentz factor of $\gamma_{\beta}\sim 100$, a {\sc gw} amplitude $h_{\mathrm{lc}} = 10^{-3}R_{\mathrm{lc}}/r$ and $R_{\mathrm{max}}\sim 10^{14}\ \mathrm{m}$ we find that the total energy transferred from the {\sc gw} to the {\sc msw} in the plasma is
\bq\label{eq::mswpower}
W_{\mathrm{tot}}  (\kappa=1)= \int^{R_{\mathrm{max}}}_{R_{\mathrm{max}} - \mathcal{H}}  W_{M}(r) \pi r^{2}\ dr \sim 10^{35}\ \mathrm{J}\ .
\eq

\subsubsection{$\kappa = \tfrac{1}{2}$}
In Paper~III we made a numerical estimate of the power density in inverse Compton scattered radiation by neglecting the magnetic field in the scattering process and using \citet{ryblight} and \eqref{eq::beamed}
\bq\label{eq::scatnonB}
P_\mathrm{ic}(r) \simeq c\gamma_{\beta}^{4} \beta^{2}\ \sigma_{\mathrm{T}} n(r) W_{M}(r)\ ,
\eq
with $\gamma_{\beta}$ and $n$ the Lorentz factor and particle density of the scattering particles, respectively, and as before, $W_{M}$ the energy density in incident {\sc msw}.

The power in scattered radiation not only depends on the energy density in the {\sc mhd} waves $W_{M}(r)$ but also on the density of scattering particles $n(r)$, and the parametrised distance dependence of the {\sc ic} power density follows from \eqref{eq::parameter} as
\bq\label{eq::icpower}
P_\mathrm{ic} \propto x^{-4\kappa}\ .
\eq
To obtain a volume integrated power we assumed in Paper~III a collimation parameter of 
$\kappa = 1/2$ in which case the spatial dependence in \eqref{eq::icpower} is the same $1/r^{2}$ decay as for the {\sc msw} energy density for the $\kappa=1$ collimation in \eqref{eq::wkappa}. However, we overestimated the total scattered power by integrating over a spherical shell $\propto r^{3}$. From \eqref{eq::vol} and \eqref{eq::icpower} we find that for $\kappa = 1/2$ the volume integrated power is a constant and that it increases with distance only for $2 \kappa > 1$. Hence, we will next consider a collimation of $\kappa = 1/3$.

\subsubsection{$\kappa = \tfrac{1}{3}$} 
Clearly, for $\kappa = 1/3$ the magnetic field and hence the {\sc msw} energy density remain much higher throughout the wind, but the decrease in integration volume is also significant as can be seen from \eqref{eq::vol} and \eqref{eq::icpower}.
The total energy deposited in {\sc msw} by the {\sc gw} for the same parameters as in Section~\ref{sec::kappa1} and over the last scale-height, which is now $\mathcal{H} = 7 r$ from \eqref{eq::scaleheight} and with $r = \sqrt{x^{2}+ R^{2}} \simeq r$ is 
\bq\label{eq::mswpowerkappa13}
W_{\mathrm{tot}} (\kappa=\tfrac{1}{3})= \int^{R_{\mathrm{max}}}_{R_{\mathrm{max}}/8}  W_{M}(r) \pi r^{\frac{2}{3}}\ dr \sim 10^{28}\ \mathrm{J}\ .
\eq
Note that from \eqref{eq::parameter} and \eqref{eq::brs} it is clear that the volume integrated energy in the {\sc msw} {\em always} increases linearly with distance, which is a consequence of flux conservation and the discussion leading to \eqref{eq::brs} (such that $h \mathcal{H}= \mathrm{const.}$).

If we repeat the estimate in \citet{moortgatIII} and integrate \eqref{eq::scatnonB} 
but now for $\kappa =1/3$ such that 
\bq
n(r) = \frac{4 M \epsilon_{0} \Omega_{b} B_{\mathrm{lc}}}{e} \left(\frac{R_{\mathrm{lc}}}{r}\right)^{\tfrac{2}{3}} \ ,
\eq derived from the Goldreich-Julian density in terms of the multiplicity $M\sim 10^{3}$ and orbital frequency $\Omega_{b}$, and
\bq\label{eq::totalic}
P_{\mathrm{ic, tot}} = c\gamma_{\beta}^{4} \ \sigma_{\mathrm{T}}\int^{R_{\mathrm{max}}}_{R_{\mathrm{max}}/8}  W_{M}(r) n(r) \pi r^{\frac{2}{3}}\ dr \sim 10^{36}\ \mathrm{W}\ ,
\eq
which for a binary merger close enough for a {\sc ligo} detection, e.g.~in the Virgo cluster at a distance of $\sim 1$~Gpc, would yield a flux on earth observable by {\sc lofar}.
The duration of this kind of transient signal is relatively short: $\Delta t = \mathcal{H}/(2 \gamma_{\beta}^2 c) \approx 15$ sec, but still much longer than the cosmic ray events that {\sc lofar} will also detect.

Unfortunately, we have learned in this paper that a more careful analysis including the polarisation of the incident and scattered wave modes with respect to the strong ambient magnetic field leads to a strong suppression of the emitted power in the scattered radio waves. 
We can find a relation similar to \eqref{eq::totalic} from \eqref{eq::energyevol} by considering mono-energetic particles with, say, a Lorentz factor $\gamma_{\beta}(\beta)\sim 10^{2}$ and distribution function $g(p_{\|}) =  \delta(p_{\|} - m \beta) n/(2\pi m)$. To evaluate the integral over wavelengths in \eqref{eq::energyevol} we assume that the incident {\sc gw} and {\sc mhd} waves are mono-chromatic with $k=k_\mathrm{gw}$.
In comparing \eqref{eq::totalic} with \eqref{eq::energyevol}, however, the latter has an additional factor $w$, the scattering probability, which depends on the small value of \eqref{eq::suppress}.

\begin{figure}
\resizebox{\hsize}{!}{\includegraphics{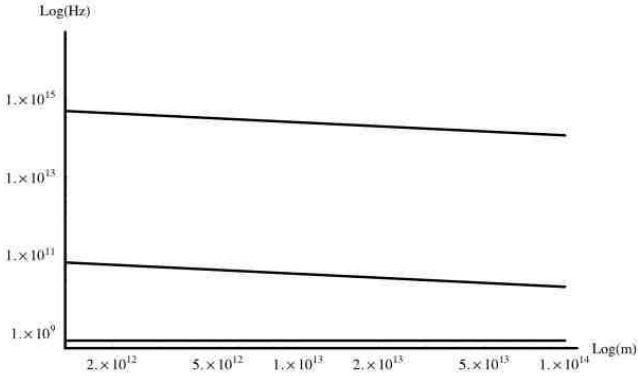}}
\caption{\label{fig::freqs}Relevant frequencies as a function of distance. Top: (non-relativistic) gyro-frequency; Middle: (non-relativistic) plasma frequency; Bottom: characteristic {\sc lofar} frequency of $100$~MHz.}
\end{figure}

In Fig.~\ref{fig::freqs} we plot the (non-relativistic) gyro-frequency $\Omega_{c}$ for a pulsar strength magnetic field ($B_{\star} = 10^{8}$~T) as a function of the distance in the wind, together with the plasma frequency (middle line) and for comparison a characteristic {\sc lofar} frequency of $100$~MHz (bottom line). Although the relativistic gyro- and plasma-frequencies are suppressed by the Lorentz factor, it is clear from Fig.~\ref{fig::freqs} that even for $\gamma \sim 10^{2}$--$10^{7}$, the gyro-frequency greatly exceeds the {\sc msw} frequency and even the frequency of the scattered waves everywhere in the wind.

%%%%%%%%%%%%
% which 
%for the magnetar magnetic field, collimated in the jet by $\kappa = 1/3$, leads to $\gamma \omega_{\mathrm{s}}/\Omega_{c} \sim 10^{-7}$ at our maximum distance $R_\mathrm{max}= 10^{14}$~m and for $\omega_{s}=100$~MHz, $\gamma\sim 100$. If instead we consider a typical pulsar surface field strength of $B_{\star}\sim 10^{8}$~T, we find that at $R_\mathrm{max}$  the scattered radio frequencies in the {\sc lofar} band are still smaller than the relativistic gyro-frequency by $\gamma \omega_{\mathrm{s}}/\Omega_{c} \sim 10^{-3}$.

\subsection{Synchrotron radiation}
In the discussion so far we have assumed that the ambient magnetic field in the wind is sufficiently strong that any transverse momentum is radiated on a time-scale short to that of the other processes considered.  This allowed us to treat the particle distribution function as strictly one-dimensional. 
However, if the magnetic energy density drops sufficiently, the small gyro-radius approximation in Section~\ref{sec::current1} is not valid anymore. Particles acquire a pitch-angle and synchrotron radiation can become important. 

We calculate whether, as the magnetic field drops to its minimum strength, either the radius of curvature of the magnetic field or the {\sc msw} wavelength could become comparable to the maximum gyro-radius $r_{c}(R_{\mathrm{max}})$ of particles that acquire some transverse momentum. In the former case, the adiabatic invariant of the particle motion, the magnetic flux through the particles orbit $\pi r_{c}^{2} B^{0}=\mathrm{const.}$ could be violated. 

However, we find that even when the particles orbit the magnetic field with relativistic velocity $v_{\bot} \sim v_{\|} \simeq c$ and corresponding Lorentz factor  $\gamma_{c}\sim 100$, we find
\bq
r_{c}(R_{\mathrm{max}}) = \frac{\gamma_{c} c}{\Omega_{c}} \simeq 0.3\ 10^{-3}\mathrm{m} \brak{\frac{\gamma_{c}}{100}}\brak{\frac{B_{\star}}{10^{8}\ \mathrm{T}}}^{-1}\ ,
\eq
which is clearly much smaller than the curvature radius of the magnetic field (which is proportional to $\mathcal{H}^{1/3}$) and the {\sc msw} wavelength $\lambda_{\mathrm{msw}} = 2 \pi c/\omega_{\mathrm{gw}} = 150\ \mathrm{km}$.

The smallness of $r_{c}$ is partly due to the rapid rotation of the source, causing the magnetic dipole field to drop only by a factor $\sim 125$ from the surface to the light-cylinder. In the wind, the field remains strong because of the force-free condition combined with the collimation. For $\kappa=1/3$, it only decreases by another factor $\sim 10^{3}$ in the entire wind and at $R_{\mathrm{max}}=10^{14}\ \mathrm{m}$ the field is still $B_{\phi}^{(0)}(R_{\mathrm{max}}) \sim 0.7 \ 10^{3}\ \mathrm{T}$. Even for $\kappa=1$, though, we still have $B_{\phi}^{(0)}(R_{\mathrm{max}}) \sim 0.4\ 10^{-3}\ \mathrm{T}$ and $r_{c}(R_{\mathrm{max}}) \sim 0.4\ \mathrm{km}$ (and now the radius of curvature of the magnetic field is $\propto \mathcal{H} \propto R_\mathrm{max}$).

The comparison of $r_{c}$ to the wavelength of the magnetosonic waves is related to the factor \eqref{eq::suppress} suppressing  the inverse-Compton scattering
\bq\label{eq::spp}
\frac{r_{c}(r)}{\lambda_\mathrm{msw}} = \frac{1}{2\pi} \frac{\gamma_{c}\omega_\mathrm{gw}}{\Omega_{c}(r)}\ .
\eq
%and we find that for $\omega_{s}/(2\pi) \simeq 100$ MHz, \eqref{eq::spp} would only become of order unity for $r\sim 10^{7}\ \mathrm{pc}$.
Consequently, if the electrons and positrons are initially confined to the magnetic field lines, it is unlikely that they will acquire transverse momentum and emit synchrotron radiation anywhere in the wind.

\section{Discussion}\label{sec::disc}
Since a strong magnetic field is essential to have efficient interaction between {\sc gw} and {\sc msw}, and millisecond rotation (with a correspondingly tight light-cylinder at $R_{\mathrm{lc}}\sim 50$~km) is necessary to have sufficiently large frequencies for the {\sc msw} and the scattered waves, $\gamma\omega/\Omega$ will be small throughout the wind.
For the range of parameters studied here, frequencies in the {\sc lofar} band never exceed the local relativistic gyro-frequency within the entire extent of the wind.

Furthermore, we know from Paper I and Paper II that the {\sc gw} most efficiently excite purely transverse magneto-acoustic waves, so the scattering channel from a (quasi-)longitudinal to another (quasi-)longitudinal mode, which doesn't suffer from cyclotron suppression, is not a viable alternative. 
The reason that scattering is suppressed is that the incident {\sc msw} does not have an electric field component along the magnetic field, and since the electrons and positrons are assumed to be constrained along the field lines they are unable to oscillate in the applied electric field perturbation.

Given the large amount of energy transferred from the {\sc gw} to the {\sc msw} in the plasma \eqref{eq::mswpower} it is very likely that some radiation mechanism will produce observable radiation in a higher frequency band. The analysis presented here proves that the simplest possibilities, synchrotron radiation, and inverse Compton scattering of the {\sc gw} induced {\sc msw} into low-frequency radio waves are ineffective. We have yet to find a more viable alternative.

\section{Conclusions}\label{sec::conclusions}
In this paper we have studied scattering processes in an essentially one-dimensional, intrinsically relativistic, magnetised plasma in some detail with the aim of investigating whether very low-frequency magnetosonic waves excited by a passing {\sc gw} can scatter on the relativistic electrons and positrons in the wind to produce radio emission in the frequency regime observable with {\sc lofar}. To treat the highly relativistic system properly, we used the covariant and gauge independent theory of plasma dynamics developed in \citet{gedalin98, melrose99, melrosegedalin99,melrose01, luomelrose, melrose02}. This approach also allows clear comparison with our general relativistic {\sc mhd} treatment in Paper I and Paper II of the propagation of gravitational waves through a magnetised plasma.  By specifying to the temporal gauge, our results are translated to more familiar vector expressions. To our knowledge, the explicit expressions for the scattering probabilities for inverse Compton scattering in a strongly magnetised, force-free and intrinsically relativistic plasma jet have not been derived before in the literature. These results are useful as well in the theory of ordinary pulsar winds and the jets of X-ray binaries and active galactic nuclei.

We assumed that the magneto-acoustic waves are excited in a cold, essentially mono-energetic plasma that satisfies the ideal {\sc mhd} condition. As they propagate outwards these waves are assumed to encounter either high energy tails of the particle distribution or, in general, a region with an intrinsically relativistic particle distribution on which they scatter. Since the plasma is strongly magnetised, all momentum transverse to the magnetic field is synchrotron radiated on a short time-scale, and as a result the charged particles are strictly confined to the magnetic field lines. In evaluating the response of the plasma, we therefore average over the relativistic one-dimensional 
J{\"u}ttner-Synge distribution function.

To simplify our results we restrict ourselves to the long wavelength regime with $\omega \ll \Omega_{c}$, which is justified in Section~\ref{sec::discussion}. Since the incident magnetosonic waves excited by the {\sc gw} are strictly transverse to the magnetic field, the scattering probability for all possible scattered modes is found to be suppressed by a factor $(\gamma\omega/\Omega_{c})^{4}$. Thus, inverse Compton scattering of the kHz {\sc msw} to $\sim 100$~MHz radio waves does not seem to be a preferred radiation mechanism for the energy dumped in the plasma by the {\sc gw}. 

The covariant formalism for relativistic plasma dynamics that we have used was developed primarily to study emission mechanisms in pulsar magnetospheres where the dipolar field lines point out almost radially from the polar caps, the charged particles flow out along these field lines and the most interesting scattering processes usually have both the incident and scattered wave vectors almost parallel to the magnetic field. The situation that we have studied is the opposite: in the force-free wind well outside the light cylinder, the magnetic field is dominated by the toroidal component and is essentially perpendicular to the incident and scattered modes of interest (in the observer frame). For a millisecond pulsar and a slightly collimated wind, the magnetic field is very strong throughout the wind and the particles move along the field lines. Note, however, that the particles have relativistic velocities both in the radial direction and along the magnetic field lines and in the laboratory frame appear to be moving outwards like `beads on a spiralling wire' predominantly in the radial direction. Consequently, the angle between both the incident and scattered wave vectors and the velocity of the particles along the magnetic field can be quite small in the lab frame (due to relativistic beaming), while they are almost at right angles in the (average) rest frame.

%%%%%%%%%%%%% The Bibliography %%%%%%%
%\bibliography{scatter} 

%%%%%%%%%%%%%%%%%%%%%%%%%%%%%%

\end{document}